\documentclass[12pt]{article}
\newcommand{\Comment}[1]{{}}
\usepackage{amsfonts,amsthm,amsmath,amssymb,slashed}
\usepackage[textwidth = 430 pt, textheight = 630 pt]{geometry}

\Comment{\usepackage{color}
\definecolor{MyDarkBlue}{rgb}{0.15,0.15,0.45}
\usepackage[linktocpage=true]{hyperref}
\hypersetup{
colorlinks=true,
citecolor=MyDarkBlue,
linkcolor=MyDarkBlue,
urlcolor=MyDarkBlue,
pdfauthor={Prieslei Goulart and Horatiu Nastase},
pdftitle={Massive ABJM and black hole entropy in the presence of field strength coupling to curvature},
pdfsubject={hep-th}
}

\usepackage[numbers,sort&compress]{natbib}
\usepackage{hypernat}}
\usepackage{graphicx}

\newcommand\ignore[1]{}
\def\one{{\,\hbox{1\kern-.8mm l}}}

\def\a{\alpha}\def\b{\beta}

\def\d{\partial}

\newcommand{\Cset}{{\,\,{{{^{_{\pmb{\mid}}}}\kern-.45em{\mathrm C}}}}}

\newcommand{\be}{\begin{equation}}
\newcommand{\bea}{\begin{eqnarray}}

\newcommand{\ee}{\end{equation}}
\newcommand{\eea}{\end{eqnarray}}

\begin{document}

\renewcommand{\thefootnote}{\fnsymbol{footnote}}

\makeatletter
\@addtoreset{equation}{section}
\makeatother
\renewcommand{\theequation}{\thesection.\arabic{equation}}

\rightline{}
\rightline{}
   \vspace{1.8truecm}


\vspace{10pt}


\begin{center}
{\LARGE \bf{\sc Massive ABJM and black hole entropy in the presence of field strength coupling to curvature}}
\end{center}
 \vspace{1truecm}
\thispagestyle{empty} \centerline{
 {\large \bf {\sc Prieslei Goulart}}\footnote{E-mail address: \Comment{\href{mailto:prieslei@ift.unesp.br}}{\tt prieslei@ift.unesp.br}}
 {\bf{\sc and}}
{\large \bf {\sc Horatiu Nastase}}\footnote{E-mail address: \Comment{\href{mailto:nastase@ift.unesp.br}}{\tt
    nastase@ift.unesp.br}}
                                                           }

\vspace{1cm}

\vspace{.8cm}
\centerline{{\it 
Instituto de F\'{i}sica Te\'{o}rica, UNESP-Universidade Estadual Paulista}} \centerline{{\it
R. Dr. Bento T. Ferraz 271, Bl. II, Sao Paulo 01140-070, SP, Brazil}}

\vspace{1.0truecm}

\thispagestyle{empty}

\centerline{\sc Abstract}

\vspace{.4truecm}

\begin{center}
\begin{minipage}[c]{380pt}
{\noindent Assuming that the near horizon geometry of the black hole solution of the gravity dual to the ABJM model, in the presence of a coupling 
between the Weyl tensor and the field strength, is $AdS_{2}\times S^{2}$, we compute Sen's entropy function for this theory. By 
extremizing the entropy function we write a formula for the entropy of the black hole, and then we compute the same entropy using 
Wald's formula and show that the results are the same. In this way we generalize the calculation of black hole entropy to cases of 
curvature coupling to the field strength, including at first order, 
and we also show how to calculate the black hole entropy when the black hole solution is unknown, 
from just a few simple assumptions about the horizon. 
}
\end{minipage}
\end{center}

\vspace{.5cm}

\setcounter{page}{0}
\setcounter{tocdepth}{2}

\newpage

\renewcommand{\thefootnote}{\arabic{footnote}}
\setcounter{footnote}{0}

\linespread{1.1}
\parskip 4pt


\section{Introduction}

Black hole solutions of supergravity are often times difficult to calculate. But in the context of AdS/CFT, they are very useful, since they determine the 
thermodynamics (at nonzero temperature) of the field theory dual to the background in which the black hole lives. The question arises then, can one 
calculate properties of the black hole which depend only on the horizon, without knowing the full solution?

One such example is provided by the attractor mechanism, that says that independent of the values of the fields at infinity, the values of
the scalars at the horizon of an extremal (be it supersymmetric or non-supersymmetric) black hole are found from the attractor equations. 
Those equations can be found from Sen's entropy function formalism \cite{Sen:2005wa} (see also \cite{Sen:2007qy}), by extremizing the entropy 
function. The entropy function at the extremum then gives the entropy of the extremal black hole. For the case of $AdS_5$ (relevant for 
the usual AdS/CFT) with higher derivative gravity, the formalism was shown to work in \cite{Astefanesei:2007vh} (see also earlier work in 
\cite{Goldstein:2005hq}).

Sometimes one does not even know a supergravity description of the background in which the extremal black hole lives. One such example 
is the gravity dual of the massive deformation \cite{Gomis:2008vc} (see also \cite{Terashima:2008sy})
of the ABJM model \cite{Aharony:2008ug}, where the gravity dual is 
only known in an implicit form \cite{Auzzi:2009es} (see also \cite{Mohammed:2010eb} for an explicit, but un-backreacted dual to massive ABJM). 
In \cite{Lopez-Arcos:2013uga} it was shown that nevertheless we can calculate some things about
the near-horizon geometry of the black holes inside this gravity dual, and using the membrane paradigm, we can derive the electric conductivity of the 
massive ABJM field theory. 

The question arises then, can we still calculate the thermodynamics of extremal black hole in these backgrounds? In this paper we answer this question
in the affirmative, for the example above, of the gravity dual to the massive ABJM model. Based on the analysis in \cite{Lopez-Arcos:2013uga}, 
a certain supergravity action, with a coupling of the Weyl tensor to the gauge field strengths,
 can be used to describe the near-horizon geometry, and we can use it to apply the entropy function formalism. 
But one needs a verification that the entropy function formalism, which to our knowledge has not been used in the presence of such a coupling
\footnote{Corrections to the entropy coming from higher order terms in the gravity action, going like contractions of the Riemann tensor squared, 
sometimes multiplied by field strengths, have been considered before. The relevant literature up to 2005 is reviewed in 
\cite{Mohaupt:2005jd}, and the relevant calculations were based on the work in \cite{Dabholkar:2004yr}. An application for the entropy function formalism
and AdS/CFT was considered in \cite{Sen:2008vm}.
However, we consider the case of 
a correction at linear (leading) order in the Riemann tensor, times field strengths. Also, the checks of the entropy function formalism in these works
were somewhat implicit (see \cite{Sahoo:2006rp} for instance), in this paper we present more explicit calculations.}, still works.

It turns out that the verification involves a new application of the Wald entropy formula. Indeed, in the case of the coupling of the curvature 
(Riemann tensor and its contractions) with gauge field strengths, it is not completely clear that the entropy formula, initially used for gravity with 
higher order terms in the Riemann tensor alone, can still be used. We find however that we obtain the same result in both formalisms, thus 
providing evidence for the corectness of both. Note that the case we are considering here, of a term in the Lagrangean with a 
single Riemann tensor contracted with field
strengths, is qualitatively different than the case with higher order corrections in the Riemann tensor, times possible field strengths, since now 
the gravitational Wald entropy will contain a contribution that is independent of the curvature of space.

The paper is organized as follows. In section 2 we describe Sen's entropy function formalism and the attractor equations, and in section 3 
we apply it for the case of the gravity dual to the massive ABJM model. In section 4 we calculate the explicit thermodynamics for a particular case 
for the charges. In section 5 we compute the entropy through the use of Wald's formula. In section 6 we show how to generalize the analysis to other 
cases of interest and in section 7 we conclude.
Appendix A describes our contraction conventions for the curvature coupling, and in Appendix B we analyze as a toy model the case with 
curvature coupling, but no scalars.

\section{Entropy function and attractor equations}

In order to compute the Sen's entropy function we assume that the spherically symmetric extremal black hole solution has the near 
horizon geometry given by\footnote{An extremal black hole is believed to have the $SO(2,1)\times SO(d-1)$ symmetry of $AdS_2\times S^{d-2}$
in the near horizon \cite{Sen:2007qy}. This has been proven in 4 and 5 dimensions \cite{Kunduri:2007vf}.}
\be 
ds^{2}=v_{1}\left(-r^{2}dt^{2}+\frac{dr^{2}}{r^{2}}\right)+v_{2}(d\theta^{2}+\sin^{2} \theta d\phi^{2}), \label{one} 
\ee
where the constants $v_{1}$ and $v_{2}$ are the $AdS_{2}$ radius and the $S^{2}$ radius respectively. The scalar and vector 
fields are constants for this geometry and are written as
\be 
\phi_{s}=u_{s}, \,\,\, F^{(A)}_{rt}=e_{A}, \,\,\, F_{\theta\phi}^{(A)}=\frac{p_{A}}{4\pi}\sin \theta, 
\ee
where ${e_{A}}$ and ${p_{A}}$ are related to the integrals of the magnetic and electric fluxes, which are in turn related to the electric and 
magnetic charges respectively. The metric (\ref{one}) has the $SO(2,1)\times SO(3)$ symmetry of $AdS_{2}\times S^{2}$. Define a 
function $f(u_{s}, v_{i}, e_{A}, p_{A})$ as the Lagrangian density $\sqrt{- \det g}{\mathcal{L}}$ evaluated for the near horizon geometry 
(\ref{one}) and integrated over the angular variables, 
\be 
f(u_{s}, v_{i}, e_{A}, p_{A})=\int d\theta d\phi \sqrt{- \det g}{\mathcal{L}}. 
\ee
We extremize this function with respect to $u_{s}$, $v_{i}$ and $e_{A}$ by
\be 
\frac{\partial f}{\partial u_{s}}=0, \,\,\,\ \frac{\partial f}{\partial v_{i}}=0, 
\ee
where these first two equations are the equations of motion for the scalar and the metric respectively. Next, one defines
the entropy function,
\be 
{\mathcal{E}}(\vec{u},\vec{v},\vec{e},\vec{q},\vec{p})\equiv 2\pi[e_{i}q_{i}-f(\vec{u},\vec{v},\vec{e},\vec{p})]. \label{entrfct}
\ee
The equations that extremize the entropy function are
\be 
\frac{\partial {\mathcal{E}}}{\partial u_{s}}=0, \,\,\, \frac{\partial {\mathcal{E}}}{\partial v_{1}}=0, \,\,\, \frac{\partial {\mathcal{E}}}{\partial v_{2}}=0, 
\,\,\, \frac{\partial {\mathcal{E}}}{\partial e_{A}}=0\;, \label{attracteqs}
\ee
and are called the attractor equations. One can show that, at the extremum, this new function equals the entropy of the black hole
\be 
S_{BH}={\mathcal{E}}(\vec{u},\vec{v},\vec{e},\vec{q},\vec{p}), 
\ee
justifying the name.

\section{Entropy function for the gravity dual to the massive ABJM model}

The 4 dimensional action used to describe the gravity dual of the massively deformed ABJM model \cite{Lopez-Arcos:2013uga} is\footnote{A note
on dimensions. With our conventions, $[e_A]=[p_A]=[q_A]=0$, $[v_i]=-2$, $[L]=-1$, $[r]=[t]=0$, $[\sqrt{-g}]=-4$, $g^{\mu\nu}=2$, $[f]=0$, $[{\cal L}]=4$, 
$[T_{AB}]=[X_A]=0$, $[g_4]=[\gamma]=0$.}
\begin{eqnarray} 
I=\int d^{4}x \sqrt{-g}\left[\frac{1}{16\pi G}\left(R-\frac{1}{2}m^{2}\left((Tr T)^{2}-2Tr(T^{2})\right)
-\frac{1}{4}Tr (\partial_{\mu} T^{-1} \partial^{\mu}T)\right)\right.\nonumber \\ \left. +\frac{1}{g_{4}^{2}}\left(-\frac{1}{4}
T_{AB}F^{A}_{\mu\nu}{F^{B}}^{\mu\nu}+\gamma L^{2}C_{\mu\nu\rho\sigma}{F^{A}}^{\mu\rho}{F^{A}}^{\nu\sigma}\right)\right], 
\label{acion}
\end{eqnarray}
where $ T_{AB}\equiv X_{A}\delta_{AB}$, and $A=0,...,7$. More precisely, this action can be used for the description of the near-horizon 
geometry of the black hole in this background, and it arises partly from a reduction of 11 dimensional supergravity down to 4 dimensions, 
on the $AdS_4\times S^7$ background, leading to gauged 4 dimensional supergravity. The effect of the nonlinear reduction in string theory 
down to 4 dimensions was argued in \cite{Myers:2010pk} to be encoded (after field redefinitions) into the Weyl coupling to field strength.

One thing is worth noting in this action: we have used a contraction 
between the Weyl tensor and the field strength that is slightly different from the one used in \cite{Myers:2010pk}, namely 
$C_{\mu\nu\rho\sigma}{F^{A}}^{\mu\nu}{F^{A}}^{\rho\sigma}$, which differs by a factor of 2 from that one. See Appendix A for more details.
The quantities of interest for us are the Riemann tensor
\begin{eqnarray}
R_{\alpha\beta\gamma\delta}=-v_{1}^{-1}(g_{\alpha\gamma}g_{\beta\delta}-g_{\alpha\delta}g_{\beta\gamma}), \,\,\, 
\alpha, \beta, \gamma, \delta=r, t \nonumber \\
R_{mnpq}=v_{2}^{-1}(g_{mp}g_{nq}-g_{mq}g_{np}), \,\,\, m, n, p, q=\theta, \phi,
\end{eqnarray}
and the Weyl tensor, written in components and with indices up for convenience,
\be 
C^{rtrt}=\frac{1}{3}\left(\frac{1}{v_{1}^{3}}-\frac{1}{v_{1}^{2}v_{2}}\right), \,\, C^{\theta\phi\theta\phi}=\frac{1}{3\sin^{2}\theta}\left(-\frac{1}{v_{1}v_{2}^{2}}+\frac{1}{v_{2}^{3}}\right). 
\ee
The near horizon gauge fields and the potential term of (\ref{acion}) can be written as
\be 
{F^{A}}_{rt}=e^{A}, \,\, {F^{A}}_{\theta\phi}=\frac{p^{A}\sin{\theta}}{4\pi}, \,\, V(X_A)=\frac{m^{2}}{2(16\pi G)}\left(-\sum_{A}X_{A}^{2}
+2\sum_{A<B}X_{A}X_{B}\right).\ee
Integrating the Lagrangian over the angular variables we have
\begin{eqnarray}
f=4\pi v_{1}v_{2}\left\lbrace \frac{1}{16\pi G}\left( -\frac{2}{v_{1}}+\frac{2}{v_{2}} \right)+\frac{2}{g_{4}^{2}}\left[\frac{1}{4v_{1}^{2}}X_{A}
+\frac{\gamma L^{2}}{3}\left(\frac{1}{v_{1}^{3}}-\frac{1}{v_{1}^{2}v_{2}} \right) \right]e^{A}e^{A} \right. \nonumber \\
+ \left. \frac{2}{g_{4}^{2}}\left[ -\frac{1}{4v_{2}^{2}}X_{A}+\frac{\gamma L^{2}}{3}\left(\frac{1}{v_{2}^{3}}
-\frac{1}{v_{1}v_{2}^{2}} \right)\right]\frac{p^{A}p^{A}}{(4\pi)^{2}} -V(X_{A})\right\rbrace . 
\end{eqnarray}
Using the definition of the entropy function(\ref{entrfct}) we get
\begin{eqnarray}
{\mathcal{E}}=2\pi\left\lbrace Q_{A}e^{A}-4\pi \left\lbrace  \frac{2(v_{1}-v_{2})}{16\pi G} +\frac{2}{g_{4}^{2}}
\left[\frac{1}{4}\frac{v_{2}}{v_{1}}X_{A}+\frac{\gamma L^{2}}{3}\left(\frac{v_{2}}{v_{1}^{2}}-\frac{1}{v_{1}} \right)
 \right]e^{A}e^{A} \right.\right. \nonumber \\
+\left.\left. \frac{2}{g_{4}^{2}}\left[ -\frac{1}{4}\frac{v_{1}}{v_{2}}X_{A}+\frac{\gamma L^{2}}{3}\left(\frac{v_{1}}{v_{2}^{2}}
-\frac{1}{v_{2}} \right)\right]\frac{p^{A}p^{A}}{(4\pi)^{2}}-v_{1}v_{2}V(X_{A}) \right\rbrace \right\rbrace . 
\end{eqnarray}
This is the entropy function for the massive dual ABJM model, when the near horizon geometry is given by (\ref{one}), 
and we can obtain the black hole entropy from it. In order to extremize the entropy function we calculate the attractor equations 
(\ref{attracteqs}),
\be 
Q_{A}=\frac{16\pi}{g_{4}^{2}}\left[\frac{1}{4}\frac{v_{2}}{v_{1}}X_{A}+\frac{\gamma L^{2}}{3}\left(\frac{v_{2}}{v_{1}^{2}}-\frac{1}{v_{1}} 
\right) \right]e^{A}, \label{QA}
\ee
\begin{eqnarray}
\frac{\partial{\mathcal{E}}}{\partial v_{1}}=-8\pi^{2} \left\lbrace \frac{2}{16\pi G} +\frac{2}{g_{4}^{2}}\left[-\frac{1}{4}\frac{v_{2}}{v_{1}^{2}}X_{A}
+\frac{\gamma L^{2}}{3}\left(-\frac{2v_{2}}{v_{1}^{3}}+\frac{1}{v_{1}^{2}} \right) \right]e^{A}e^{A} \right. \nonumber \\
+ \left. \frac{2}{g_{4}^{2}}\left[ -\frac{1}{4}\frac{1}{v_{2}}X_{A}+\frac{\gamma L^{2}}{3}\frac{1}{v_{2}^{2}} \right]\frac{p^{A}p^{A}}{(4\pi)^{2}}
-v_{2}V(X_{A}) \right\rbrace =0, \label{a1} 
\end{eqnarray}
\begin{eqnarray}
\frac{\partial{\mathcal{E}}}{\partial v_{2}}=-8\pi^{2} \left\lbrace -\frac{2}{16\pi G} +\frac{2}{g_{4}^{2}}\left[\frac{1}{4}\frac{1}{v_{1}}X_{A}
+\frac{\gamma L^{2}}{3}\left(\frac{1}{v_{1}^{2}}\right) \right]e^{A}e^{A} \right. \nonumber \\
+\left. \frac{2}{g_{4}^{2}}\left[ \frac{1}{4}\frac{v_{1}}{v_{2}^{2}}X_{A}+\frac{\gamma L^{2}}{3}\left(\frac{1}{v_{2}^{2}}-\frac{2v_{1}}{v_{2}^{3}} 
\right)\right]\frac{p^{A}p^{A}}{(4\pi)^{2}}-v_{1}V(X_{A}) \right\rbrace =0. \label{a2}
\end{eqnarray}
Using the combination $v_{1}\frac{\partial{\mathcal{E}}}{\partial v_{1}}-v_{2}\frac{\partial{\mathcal{E}}}{\partial v_{2}}=0$, and defining in the final 
expression $e^{A}=\frac{q^{A}}{4\pi}$, we find a result that will be useful later
\begin{eqnarray}
-\frac{1}{g_{4}^{2}}\frac{v_{1}v_{2}}{(4\pi)^2}\frac{1}{4}\sum_{A}X_{A}\left(\frac{q^{A}q^{A}}{v_{1}^{2}}+\frac{p^{A}p^{A}}{v_{2}^{2}}\right)=
-\frac{v_{1}+v_{2}}{32\pi G}\nonumber \\
+\frac{1}{2g_{4}^{2}}\frac{1}{(4\pi)^{2}}\frac{\gamma L^{2}}{3}\sum_{A}\left( (3v_{2}-v_{1})\frac{q^{A}q^{A}}{v_{1}^{2}}-(3v_{1}-v_{2})
\frac{p^{A}p^{A}}{v_{2}^{2}}\right).\label{intermed} 
\end{eqnarray}
Eliminating $Q_{A}$ through (\ref{QA}) from the entropy function we have
\begin{eqnarray}
{\mathcal{E}}=16\pi^{2}\left\lbrace  \frac{(v_{2}-v_{1})}{16\pi G}+ \frac{v_{1}v_{2}}{2}V(X_{A}) -\frac{1}{g_{4}^{2}}\frac{1}{4}\frac{v_{1}v_{2}}{(4\pi)^{2}}
\sum_{A}X_{A}\left(\frac{q^{A}q^{A}}{v_{1}^{2}}+\frac{p^{A}p^{A}}{v_{2}^{2}}\right)\right. \nonumber \\
+\frac{1}{g_{4}^{2}}\frac{\gamma L^{2}}{3}
\left. \frac{(v_{2}-v_{1})}{(4\pi)^{2}}\sum_{A}\left(\frac{q^{A}q^{A}}{v_{1}^{2}}+\frac{p^{A}p^{A}}{v_{2}^{2}}\right) \right\rbrace . \end{eqnarray}
Now we use the result (\ref{intermed}) and the entropy becomes
\begin{eqnarray}
{\mathcal{E}}=16\pi^{2}\left\lbrace  \frac{3v_{2} -v_{1}}{32\pi G}- \frac{1}{2g_{4}^{2}}\frac{1}{(4\pi)^{2}}\frac{\gamma L^{2}}{3}
(v_{1}+v_{2})\sum_{A}\left(\frac{q^{A}q^{A}}{v_{1}^{2}}-\frac{p^{A}p^{A}}{v_{2}^{2}}\right)\right. \nonumber \\
\left. +\frac{v_{1}v_{2}}{2}V(X_{A}) \right\rbrace .  
\end{eqnarray}
The potential can be eliminated from this expression by using $\frac{\partial {\mathcal{E}}}{\partial v_{1}}\cdot \frac{1}{v_{2}}
+\frac{\partial {\mathcal {E}}}{\partial v_{2}}\cdot \frac{1}{v_{1}}=0$, i.e.
\be 
V(X_{A})=\frac{1}{16\pi G}\left(\frac{1}{v_{2}}-\frac{1}{v_{1}}\right)+\alpha\left(-\frac{1}{v_{1}}+\frac{1}{v_{2}}\right)\sum_{A}
\left(\frac{q^{A}q^{A}}{v_{1}^{2}}-\frac{p^{A}p^{A}}{v_{2}^{2}}\right), \label{intermed2}
\ee
where we have defined the constant
\be 
\alpha \equiv \frac{1}{g_{4}^{2}}\frac{1}{(4\pi)^{2}}\frac{\gamma L^{2}}{3}. 
\ee
Replacing the potential in ${\mathcal{E}}$ we obtain finally the entropy of the black hole
\be 
{\mathcal{E}}=16\pi^{2}v_{2}\left\lbrace \frac{1}{16\pi G}-\frac{\gamma L^{2}}{3 g_{4}^{2}}\frac{1}{(4\pi^{2})}\sum_{A}
\left(\frac{q^{A}q^{A}}{v_{1}^{2}}-\frac{p^{A}p^{A}}{v_{2}^{2}}\right)  \right\rbrace  . \label{bhent}
\ee
The first term gives the usual relation $area/4G$, expected from the theory without correction factors depending on the curvature. 
The second term is an extra factor that arose due to the presence of the Weyl tensor coupling in the action. The above result thus reproduces 
the usual formula for the entropy when one takes $\gamma = 0$. It is worthwhile to notice that, although there may be some symmetry 
relating the constants $q^{A}$ and $p^{A}$, the entropy is not written yet in terms of the electric charge of the black hole. In order to 
obtain that, one would have to solve the equation (\ref{QA}) and get as a result the constant $q^{A}$ writen in terms of the electric charge $Q^{A}$. 

We now show how to write the scalar fields $X_{A}$ in terms of the constants $q^{A}$, $p^{A}$, $v_{1}$ and $v_{2}$. 
Using the attractor equation for the components of the scalar field, we obtain
\begin{eqnarray} 
\frac{\partial{\mathcal{E}}}{\partial X_{A}}=-8\pi^{2}\left[\frac{2}{g_{4}^{2}}\frac{1}{4(4\pi)^2}\frac{v_{2}}{v_{1}}q^{A}q^{A}
- \frac{2}{g_{4}^{2}}\frac{1}{4(4\pi)^{2}}\frac{v_{1}}{v_{2}}p^{A}p^{A}\right.\nonumber \\
\left. -\frac{v_{1}v_{2}m^{2}}{16\pi G}\left(-X_{A}+\sum_{B\neq A}X_{B}\right)  \right]=0, 
\end{eqnarray}
which gives
\be 
X_A=\sum_{B\neq A}X_{B}-\frac{16\pi G}{2m^{2}g_4^2(4\pi)^2}\left(\frac{q^Aq^A}{v_{1}^{2}}- \frac{p^A p^A}{v_{2}^{2}}\right). 
\ee

Defining
\be 
M^{A}\equiv \frac{q^{A}q^{A}}{v_{1}^{2}}-\frac{p^{A}p^{A}}{v_{2}^{2}}, \,\,\, 
K \equiv -\frac{16\pi G}{ 2m^{2}g_4^2}\frac{1}{(4\pi)^{2}},  
\ee
we can write the scalar fields as
\be 
X_{A}=\sum_B X_B-X_A+K M^A = -\frac{K}{12}\left\lbrace \sum_{B=0}^{7} M^{B}-6M^{A}\right\rbrace . \label{scalar}
\ee
Since we have determined the scalar field as a function of the charges we can obtain the potential. We use the relations
\bea 
\sum_{A=0}^{7}X_{A}&=&\frac{K}{6}\left\lbrace 4\sum_{B=0}^{7}M^{B}-3\sum_{A=0}^{7}M^{A}\right\rbrace  ,\cr
\sum_{A=0}^{7}X_{A}^{2}&=&\frac{K^{2}}{18}\left\lbrace 4\sum_{A=0}^{7}(M^{A})^{2}-\sum_{A<B}M^{A}M^{B}\right\rbrace ,\cr
\sum_{B<D}^{7}X_{B}X_{D}&=&\frac{K^{2}}{36}\left\lbrace -14\sum_{A=0}^{7}(M^{A})^{2}-19
\sum_{B<D}^{7}M^{B}M^{D}\right\rbrace .
\eea
to calculate
\bea 
-\sum_{A=0}^{7}(X_{A})^{2}+2\sum_{A<B}^{7}X_{A}X_{B}&=&
-K^{2}\left\lbrace \sum_{A=0}^{7}(M^{A})^{2}+\sum_{B<D}^{7}M^{B}M^{D}\right\rbrace\cr
&=&-K^{2}\sum_{B\leq D}^{7}M^{B}M^{D}. 
\eea
Then the potential becomes
\bea 
V(X_{A})&=&-\frac{m^2K^2}{2(16\pi G)}
\sum_{A\leq B}M^{A}M^{B}\cr
&=&-\beta\sum_{A\leq B}\left(\frac{q^{A}q^{A}}{v_{1}^{2}}\frac{q^{B}q^{B}}{v_{1}^{2}}
+\frac{p^{A}p^{A}}{v_{2}^{2}}\frac{p^{B}p^{B}}{v_{2}^{2}}-\frac{q^{A}q^{A}}{v_{1}^{2}}\frac{p^{B}p^{B}}{v_{2}^{2}}
-\frac{p^{A}p^{A}}{v_{2}^{2}}\frac{q^{B}q^{B}}{v_{1}^{2}}    \right),\cr
&&
\eea
where 
\be 
\beta \equiv \frac{m^2K^2}{2(16\pi G)}=
\frac{16\pi G}{8(4\pi)^4m^2g_4^4}. 
\ee

\section{Explicit thermodynamics in a special case}

We now discuss solutions of the attractor equations, in order to calculate explicitly the thermodynamics of the model. 
By adding equations (\ref{a1}) and (\ref{a2}), we find
\begin{eqnarray}
\frac{2}{g_{4}^{2}(4\pi)^{2}}(v_2-v_1)\frac{X_{A}}{4}\left(\frac{q^{A}q^{A}}{v_{1}^{2}}
+\frac{p^{A}p^{A}}{v_{2}^{2}}\right)+4\alpha (v_2-v_1) \left(\frac{q^{A}q^{A}}{v_{1}^{3}}
-\frac{p^{A}p^{A}}{v_{2}^{3}} \right)\nonumber \\
 +(v_1+v_{2})V(X_{A})  =0.
\end{eqnarray}
Using the potential in  (\ref{intermed2}),  we see that a possible solution is $v_1=v_2\equiv v$, and in this section we will only consider this case.

Obtaining the solution of the equations of motion in the presence of both electric and magnetic charge is quite complicated. For simplicity, we will 
consider the case that the black hole solution has only electric charge, $p^{A}=0$, and also that the gauge fields are all the same, 
meaning $q^{A}=q$. We rewrite the 
equation of motion (\ref{intermed}), written in terms of the constants $\alpha$ and $\beta$, 
by replacing the $X_{A}$ obtained before in (\ref{scalar}), i.e.
\be 
v^3-8\alpha (16 \pi G) q^2v-(16 \pi G)\frac{4 \beta}{3}q^{4}=0. \label{cubic}
\ee
In this expression, the electric field can be found from (\ref{QA}), and it is given by
\be 
q=\left(\frac{3Qv^{2}}{32\pi^2\beta}\right)^{1/3}.\label{q} 
\ee 
Replacing $q$ in eq. (\ref{cubic}), we obtain
\be 
v^{2/3}-\frac{4}{3}\frac{v^{1/3}}{\b^{1/3}}(16\pi G) \left(\frac{3Q}{32\pi^2}\right)^{4/3}-\frac{8\a}{\b^{2/3}}(16\pi G)\left(\frac{3Q}{32\pi^2}\right)^{2/3}=0.
\ee
Then we obtain (keeping only the solution of the quadratic equation with the plus sign, the one with the minus giving, for $\a>0$, a negative 
$v$, which is unphysical) 
\bea
v&=&\frac{8(16\pi G)^2}{27\b}\left(\frac{3Q}{32\pi^2}\right)^2\left[4(16\pi G)\left(\frac{3Q}{32\pi^2}\right)^2+54\a\right.\cr
&&\left.+\left(4(16\pi G)\left(\frac{3Q}{32\pi^2}\right)^2+18\a\right)\sqrt{1+\frac{18\a}{16\pi G}\left(\frac{32\pi^2}{3Q}\right)^2}\right].
\eea
For a small perturbation in $\a$, we obtain
\be
v\simeq \frac{32}{27\b}(16\pi G)^2\left(\frac{3Q}{32\pi^2}\right)^2\left[2(16\pi G)\left(\frac{3Q}{32\pi^2}\right)^2+27\a\right].
\ee

Now we can calculate the entropy of the solution. We first substitute (\ref{q}) in (\ref{bhent}) and obtain
\bea 
{\mathcal{E}}&=& \frac{\pi v}{G}-8\a 16\pi^2\frac{q^2}{v}\cr
&=&\frac{\pi v}{G}-8\a16\pi^2 \frac{v^{1/3}}{\b^{2/3}}\left(\frac{3Q}{32\pi^2}\right)^{2/3}.
\eea
Next we substitute the solution for $v$, and get
\be
{\cal E}=\frac{8\pi(16\pi G)^2}{27\b G}\left(\frac{3Q}{32\pi^2}\right)^2\left[4(16\pi G)\left(\frac{3Q}{32\pi^2}\right)^2
\left(1+\sqrt{1+\frac{18\a}{16\pi G}\left(\frac{32\pi^2}{3Q}\right)^2}\right)+36\a\right].
\ee
For a small perturbation in $\a$, we obtain
\be 
{\mathcal{E}}\simeq \frac{3GQ^2}{4\pi \b}\left(\frac{GQ^2}{\pi^3}+64\a\right).
\ee

For the extremal black holes considered in this paper, the entropy is nonzero, but the temperature is zero, hence so will the chemical potentials.
But the chemical potentials divided by the temperature ($\mu_i/T$) are nonzero, and can be calculated as usual from 
\be
\frac{\mu_i}{T}=\frac{\d{\cal E}}{\d Q_i}\;,
\ee
where $Q_i$ are all the charges of the system. In the special case above, we can calculate $\mu/T$, obtaining
\bea
\frac{\mu}{T}&=&\frac{8\pi}{27\b G}(16\pi G)^3Q^3\left(\frac{3}{16\pi^2}\right)^4\left[1+\sqrt{1+\frac{18\a}{16\pi G}\left(\frac{32\pi^2}{3Q}\right)^2}\right]\cr
&&+\frac{16\pi\a}{3\b G}(16\pi G)^2 Q\left(\frac{3}{16\pi^2}\right)^2\left[1-\frac{1}{\sqrt{1+\frac{18\a}{16\pi G}\left(\frac{32\pi^2}{3Q}\right)^2}}\right].
\eea
For a small perturbation  in $\a$, we obtain to first order
\be
\frac{\mu}{T}\simeq\frac{3GQ}{\pi\b}\left(\frac{GQ^2}{\pi^3}+32\a\right).
\ee
The solutions with $v_{1}\neq v_2$ are very complicated, due to the third power in the electric field in (\ref{QA}), and we don't treat them here. 
By comparison with the toy model presented in appendix B, we expect that these solutions do not minimize the entropy, so don't need to be considered.

\section{Computation of the entropy by the use of Wald's formula}

In the construction of the entropy function\cite{Sen:2005wa}, Sen compared his formula with the Wald entropy formula, and showed that, at 
the extremum, the entropy function was indeed the entropy of the black hole. It is natural then to imagine that the entropy of the 
black hole calculated by using Wald's formula would agree with the previous result, and we will show that this is indeed the case here.

Wald's formula is given by
\be 
S_{W}=-2\pi \int_{\Sigma} E^{abcd}_{R}\epsilon_{ab}\epsilon_{cd}, 
\ee
where $E^{abcd}_{R}$ is the equation of motion for $R_{abcd}$, written as
\be 
E^{abcd}_{R}=\frac{\partial L}{\partial R_{abcd}}-\nabla_{a_{1}}\frac{\partial L}{\partial \nabla_{a_{1}}R_{abcd}}+...+(-1)^{m}\nabla_{\left( a_{1} ...\right.}\nabla_{ \left. a_{m} \right)}\frac{\partial L}{\partial( \nabla_{\left( a_{1}... \right.}\nabla_{ \left. a_{m} \right)}R_{abcd})},
\ee
$\epsilon_{ab}=2\nabla_{\left[a\right.}\xi_{\left. b\right]}$ is the binormal to the horizon of the black hole, the integral is taken over 
the two-surface $\Sigma$, and $\xi_{b}$ is a timelike Killing vector normalized such that $\xi_{t}\xi^{t}=-1$. For the near horizon geometry \ref{one} the Killing vector is given by
\be 
\xi_{t}\xi^{t}=g^{tt}(\xi_{t})^{2}=-1 \Rightarrow \xi_{t}=\sqrt{v_{1}}r, 
\ee
and the binormal by
\be 
\epsilon_{rt}=\nabla_{r}\xi_{t}-\nabla_{t}\xi_{r}=\partial_{r}\xi_{t}
-\frac{1}{r}\xi_{t}-\partial_{t}\xi_{r}
+\frac{1}{r}\xi_{t}=\sqrt{v_{1}}. \label{binormal}
\ee
Wald's formalism was developed to be applied to theories which are invariant under diffeomorphisms, and in order to get unambiguous 
results we must rewrite the Lagrangian in a form suitable for the computation of the entropy, following the algorithm developed in \cite{Iyer:1994ys}. 
Although one in principle needs to follow that algorithm, in general the Lagrangian would not be manifestly gauge invariant anymore. 

For theories involving Lagrangians which are also gauge invariant we must have the Lagrangian rewritten also in a 
manifestly gauge invariant form, and 
then the result would agree with the result obtained by the computation through other methods. In our case, this is achieved if we 
keep the Lagrangian in its original form, and apply Wald's formula to it. Using the definition of the Weyl tensor the relevant part 
of the Lagrangian can be cast as (see eq. (\ref{Weylcontr}))
\begin{eqnarray} 
L_{R}=\sqrt{-g}\left[\frac{1}{2\cdot 16\pi G}(g^{\mu\rho}g^{\nu\sigma}-g^{\mu\sigma}g^{\nu\rho})R_{\mu\nu\rho\sigma}
+\frac{\gamma L^{2}}{g_{4}^{2}}\left( R_{\mu\nu\rho\sigma}F^{A\mu\rho}F^{A\nu\sigma}\right. \right.\nonumber \\
\left. \left. +R_{\mu\sigma}{F^{A\mu}}_{\nu}F^{A\nu\sigma}+\frac{1}{6}R F^A_{\mu\nu}F^{A\mu\nu}\right)\right]. 
\end{eqnarray}
Computing the derivative with respect to $R_{abcd}$ and replacing $E^{abcd}_{R}$, $F^A_{rt}=\frac{q^{A}}{4\pi}$ and 
$F^A_{\theta\phi}=\frac{p^{A}\sin\theta}{4\pi}$  in the Wald formula, we get
\begin{eqnarray} 
S_{W}&=&-2\pi \int_{\Sigma}v_{1}v_{2}\sin\theta
(2g^{rr}g^{tt})\epsilon_{rt}\epsilon_{rt}
\left[\frac{1}{16\pi G}-\frac{1}{3}\frac{\gamma L^{2}}{g_{4}^{2}(4\pi)^{2}}\sum_{A} \left(\frac{q^{A}q^{A}}{v_{1}^{2}}
-\frac{p^{A}p^{A}}{v_{2}^{2}}\right)\right].\cr
&&
\end{eqnarray}
The angular integration gives $4\pi$. Replacing the value of the binormal (\ref{binormal}) and the metric elements we obtain the Wald 
entropy for the near horizon geometry of the black hole
\be 
S_{W}=16\pi^{2}v_{2} 
\left\lbrace \frac{1}{16\pi G}-\frac{1}{3}\frac{\gamma L^{2}}{g_{4}^{2}(4\pi)^{2}}\sum_{A} \left(\frac{q^{A}q^{A}}{v_{1}^{2}}
-\frac{p^{A}p^{A}}{v_{2}^{2}}\right)\right\rbrace . \label{waldent}
\ee
This result is the same as the one found by using the Sen's entropy function, the first term being the usual relation $area/4G$ and 
the second one a correction factor proportional to $\gamma$.

\section{Generalization}

We have seen that in one case, of the gravity dual to the massive ABJM model, with a coupling of the Weyl tensor to the field strength in the 
action, we can calculate the entropy of extremal black holes in two ways (via the entropy function and via the Wald formula)
and obtain the same result, which gives evidence for the correctness of the general procedure used. We can therefore propose that this 
procedure is also valid in more general cases. 

The first step of the procedure was to assume that the extremal black hole has always a near horizon geometry of $AdS_2\times S^2$ type, 
or in a general dimension $d$, of $AdS_2\times S^{d-2}$. This is a very reasonable assumption, based on evidence from many examples
and proven in 4 and 5 dimensions.
We then need to know the supergravity action that describes this near-horizon geometry. Here one would proceed case by case, but the 
starting point should be the gauged supergravity action in $d$ dimensions. In general, one would also have curvature couplings to the 
field strength, through terms like 
\be
\a R_{\mu\nu\rho\sigma}F^{A\mu\nu}F^{A\rho\sigma}+\b R_{\mu\nu}{F^{A\mu}}_\lambda F^{A\lambda\nu}+\gamma R F_{\mu\nu}^A F^{A\mu\nu}
\ee
and more, perhaps involving covariant derivatives of the Riemann tensor. Note that by field redefinitions, one can get rid of two of the above 
terms, keeping only one \cite{Myers:2009ij}, which can be chosen to be either the first, or the combination giving the Weyl tensor contraction.

For such Lagrangeans, which besides diffeomorphism invariance, also have gauge invariance, we can use Wald's entropy formula, generalized to 
this case. Note that one could in principle rewrite the Lagrangean in different ways through partial integrations, 
but we first need to write it in a manifestly gauge invariant form, 
and then we can take the derivative of the Lagrangean with respect to the Riemann tensor and its covariant derivatives, to calculate Wald's formula.

For the entropy function formulation, the generalization is straightforward, since the entropy function is defined from the integral of the Lagrangean 
as usual. Either of the two formulations, the Wald entropy, or the entropy function, can be used to calculate the extremal black hole entropy.

Note that higher order corrections considered before (see, e.g. \cite{Mohaupt:2005jd} for an early review) are at least quadratic in gravity 
(in the Riemann, or Weyl, tensor), times possible field strengths. But in these cases, the Wald formula (which is the integral of a derivative 
of ${\cal L}$ with respect to the Riemann tensor) will give correction terms to the Bekenstein-Hawking area law
that still depend on the Riemann tensor, which is the application that 
Wald had in mind. The corrections we are describing here will give a term in the Wald entropy that is independent of the Riemann tensor, so 
it was not {\em  a priori} clear that the same formula would work.

\section{Conclusions}

In this paper we have shown how to calculate the entropy of extremal black holes in cases where we don't have an explicit solution for the 
black hole or its background, and how to use Wald's formula in the cases where the we have couplings of curvature to gauge field strength. 
The example we have focused on is that of black holes in the gravity dual to the massive ABJM model. The near-horizon geometry is described 
by a supergravity action that has a coupling of one Weyl tensor to two field strength tensors, for which we can apply both Sen's entropy function 
formalism and Wald's entropy formula. The application of both in this context is new, and we found agreement, providing evidence for the 
corectness of the approaches. The approach followed here can be used in more general contexts, and we described the general procedure.
We have solved explicitly the attractor equations for the massive ABJM model, and found the thermodynamics in a particular case, $p^A=0$
and $q^A=q$.

{\bf Acknowledgements} We would like to thank Johanna Erdmenger, Jeff Murugan, Koenraad Schalm and Robert Wald for discussions.
The research of HN is supported in part by CNPQ grant 301709/2013-0
and FAPESP grant 2013/14152-7, and the research of PG is supported by FAPESP grant 2013/00140-7.

\appendix

\section{Contraction conventions for the Weyl tensor}

Note that using the first Bianchi identity for the Riemann tensor, $R_{\mu\nu\rho\sigma}+R_{\mu]\rho\sigma\nu}+R_{\mu\sigma\nu\rho}=0$, we find that 
\bea
R_{\mu\nu\rho\sigma}F^{\mu\nu}F^{\rho\sigma}&=&-R_{\mu\rho\sigma\nu}F^{\mu\nu}F^{\rho\sigma}-R_{\mu\sigma\nu\rho}F^{\mu\nu}F^{\rho\sigma}
=R_{\mu\rho\nu\sigma}F^{\mu\nu}F^{\rho\sigma}+R_{\nu\rho\mu\sigma}F^{\mu\nu}F^{\rho\sigma}\cr
&=&2R_{\mu\rho\nu\sigma}F^{\mu\nu}F^{\rho\sigma}\;,
\eea
where in the first equality we have used the Bianchi identity, in the second the symmetry of the Riemann tensor, and in the last the antisymmetry of the 
field strength. Since in general 
\be
C_{\mu\nu\rho\sigma}=R_{\mu\nu\rho\sigma}-\frac{2}{d-2}(g_{\mu[\rho}R_{\sigma]\nu}-g_{\nu[\rho}R_{\sigma]\mu})+\frac{2}{(d-1)(d-2)}R
g_{\mu[\rho}g_{\sigma]\nu}\;,
\ee
we obtain 
\bea 
C_{\mu\nu\rho\sigma}F^{\mu\nu}F^{\rho\sigma}&=&R_{\mu\nu\rho\sigma}F^{\mu\nu}F^{\rho\sigma}+
2R_{\sigma\mu}{F^{\mu}}_{\rho}F^{\rho\sigma}+\frac{1}{3}RF_{\mu\nu}F^{\mu\nu}\;, \cr
C_{\mu\nu\rho\sigma}F^{\mu\rho}F^{\nu\sigma}&=& R_{\mu\nu\rho\sigma}F^{\mu\rho}F^{\nu\sigma}+
R_{\mu\sigma}{F^{\mu}}_{\nu}F^{\nu\sigma}+\frac{1}{6}R F_{\mu\nu}F^{\mu\nu}\;,\label{Weylcontr}
\eea 
so we get
\be
C_{\mu\nu\rho\sigma}F^{\mu\nu}F^{\rho\sigma}=2C_{\mu\nu\rho\sigma}F^{\mu\rho}F^{\nu\sigma}.
\ee
We then see that with the new contraction, we need to rescale the coefficient, so 
\be 
\gamma_{Myers}=2\gamma_{ours}. 
\ee

\section{Toy model}

As a simple example of black hole entropy in the presence of field strength coupled to curvature, we consider as a toy model the theory in 
\cite{Myers:2010pk}, with only one gauge field and no scalars, but with the coupling between the Weyl tensor and the field strength, i.e.
\begin{eqnarray} 
I=\int d^{4}x \sqrt{-g}\left[\frac{1}{16\pi G}R-\frac{1}{4}
F_{\mu\nu}F^{\mu\nu}+\gamma L^{2}C_{\mu\nu\rho\sigma}F^{\mu\rho}F^{\nu\sigma}\right],\label{action2} 
\end{eqnarray}
where we have set ${g_4}^{2}=1$ for simplicity. The geometry adopted is the same as the one adopted throughout the paper. 
We write the entropy function as
\begin{eqnarray}
{\mathcal{E}}=2\pi\left\lbrace Qe-4\pi \left[  \frac{2(v_{1}-v_{2})}{16\pi G} +\frac{1}{2}\left(\frac{v_{2}}{v_{1}}
e^{2}-\frac{v_{1}}{v_{2}}\frac{p^{2}}{(4\pi)^{2}} \right) +\frac{\tilde\alpha}{2}\left( \frac{v_{2}}{v_{1}^{2}}-\frac{1}{v_{2}}\right)e^{2}\right. \right. \nonumber \\
\left. \left. +\frac{\tilde\alpha}{2}\left(\frac{v_{1}}{v_{2}^{2}}
-\frac{1}{v_{2}} \right)\frac{p^{2}}{(4\pi)^{2}} \right] \right\rbrace \;, \label{entfunction}
\end{eqnarray}
where now we have defined $\tilde\alpha \equiv (4 \gamma L^{2})/3$. Also defining $P=p/(4\pi)$, the attractor equations are
\be 
\frac{2}{16\pi G}-\frac{v_{2}}{2}\left(\frac{e^{2}}{{v_1}^{2}}+\frac{P^{2}}{{v_2}^{2}}\right)+\frac{\tilde\alpha}{2}
\left[\left(\frac{-2{v_{2}}}{{v_1}^{3}}+\frac{1}{v_{1}^{2}}\right)e^{2}+\frac{P^{2}}{{v_2}^{2}}\right]=0, \label{ateq1}
\ee
\be 
-\frac{2}{16\pi G}+\frac{{v_1}}{2}\left(\frac{e^{2}}{{v_1}^{2}}+\frac{P^{2}}{{v_2}^{2}}\right)
+\frac{\tilde\alpha}{2}\left[\frac{e^{2}}{{v_1}^{2}}+\left(\frac{-2v_{1}}{{v_2}^{3}}+\frac{1}{v_{2}^{2}}\right)P^{2}\right]=0, \label{ateq2}
\ee
\be 
Q=4\pi\left[\frac{v_{1}v_{2}+\tilde\alpha(v_{2}-v_{1})}{v_{1}^{2}}\right]e. 
\ee
Replacing $Q$ in the entropy function (\ref{entfunction}), we obtain
\begin{eqnarray}
{\mathcal{E}}=8\pi^{2}\left\lbrace \frac{-2(v_{1}-v_{2})}{16\pi G} -\frac{v_1v_2}{2}
\left(\frac{e^{2}}{v_{1}^{2}}+\frac{P^{2}}{v_{2}^{2}} \right) +\frac{\tilde\alpha}{2}\left( \frac{v_{2}}{v_{1}^{2}}-\frac{1}{v_{1}}\right)e^{2} \right. \nonumber \\
\left. -\frac{\tilde\alpha}{2}\left(\frac{v_{1}}{v_{2}^{2}}
-\frac{1}{v_{2}} \right)P^{2}  \right\rbrace  . \label{entfunction2}
\end{eqnarray}
Using the combination $v_{1}\frac{\partial{\mathcal{E}}}{\partial v_{1}}-v_{2}\frac{\partial{\mathcal{E}}}{\partial v_{2}}=0$ we can write the expression
\be 
\frac{v_{1}v_{2}}{2}\left(\frac{e^{2}}{v_{1}^{2}}+\frac{P^{2}}{v_{2}^{2}} \right)=\frac{(v_{1}+v_{2})}{16\pi G}
+\frac{\tilde\alpha}{4}\left[-3\frac{v_{2}}{v_{1}^{2}}e^{2}+3\frac{v_{1}}{v_{2}^{2}}P^{2}+\frac{e^{2}}{v_{1}}-\frac{P^{2}}{v_{2}}\right]. 
\ee
The entropy function in (\ref{entfunction2}) can be recast so that the term on the left of this expression can be recognized, 
and after replacing it by the right hand side the entropy function will be
\begin{equation}
{\mathcal{E}}=8\pi^{2}\left\lbrace \frac{3v_{2}-v_{1}}{16\pi G} -\frac{\tilde\alpha}{4}(v_{2}+v_{1})
\left( \frac{e^{2}}{v_{1}^{2}}-\frac{P^{2}}{v_{2}^{2}}\right)   \right\rbrace  . \label{entfunction3}
\end{equation}
Now we use the combination $v_{1}\frac{\partial{\mathcal{E}}}{\partial v_{1}}+v_{2}\frac{\partial{\mathcal{E}}}{\partial v_{2}}=0$ and write
\be 
-\frac{v_{2}}{16\pi G}-\frac{\tilde\alpha v_{2}}{4}\left(\frac{e^{2}}{v_{1}^{2}}-\frac{P^{2}}{v_{2}^{2}}\right)
=-\frac{v_{1}}{16\pi G}-\frac{\tilde\alpha v_{1}}{4}\left(\frac{e^{2}}{v_{1}^{2}}-\frac{P^{2}}{v_{2}^{2}}\right), \label{condition}\ee
which allows us to write eq. (\ref{entfunction3}) as
\begin{equation}
{\mathcal{E}}=16\pi^{2}v_{2}\left\lbrace \frac{1}{16\pi G} -\frac{\tilde\alpha}{4}
\left( \frac{e^{2}}{v_{1}^{2}}-\frac{P^{2}}{v_{2}^{2}}\right)   \right\rbrace  . \label{entfinal}
\end{equation}
It is very easy to adapt eq. (\ref{waldent}) to show that this is the same result obtained by applying the Wald formula. 
Adding up the attractor equations (\ref{ateq1}) and (\ref{ateq2}), we obtain
\be 
(v_{1}-v_{2})\left[\frac{1}{2}\left(\frac{e^{2}}{v_{1}^{2}}+\frac{P^{2}}{v_{2}^{2}}\right) +\tilde\alpha \left(\frac{e^{2}}{v_{1}^{3}}-\frac{P^{2}}{v_{2}^{3}}\right) \right]=0. 
\ee
There are two types of solutions. The first is $v_1=v_2$, and the second happens when the term in square brackets is equal to zero. 
For the case of $v_1=v_2\equiv v$, the attractor equations  (\ref{ateq1}) and (\ref{ateq2}) become the same, i.e.
\be 
v^{2}-\frac{(16 \pi G)}{4}(e^{2}+P^{2})v-\frac{\tilde\alpha (16 \pi G)}{4}(e^{2}-P^{2})=0,\,\,\,\, e=\frac{q}{4\pi}. 
\ee
Solving for $v$ we have
\be 
v=\frac{1}{2}\frac{(16 \pi G)}{4}(e^{2}+P^{2})\left[1\pm \sqrt{1+\frac{16\tilde\alpha}{(16 \pi G)}\frac{(e^{2}-P^{2})}{(e^{2}+P^{2})^2}}\right]. 
\ee
Replacing the solutions for $v$ and $e$ in the entropy function (\ref{entfinal}), and using $P\equiv q/4\pi$ to use the same 
notation as Sen \cite{Sen:2005wa} for the sake of comparison, the entropy of the black hole becomes
\begin{equation}
{\mathcal{E}}=\frac{q^{2}+p^{2}}{4}.
\end{equation}
This is the same entropy found by using Sen's entropy function for the extremal Reissner-Nordstrom solution in \cite{Sen:2005wa}, 
in which case the action was just the Einstein-Maxwell action. Since we have an extra term in the action, which is a coupling between 
the Weyl tensor and field strengths, we would expect to have obtained corrections for the entropy of the extremal Reissner-Nordstrom 
black hole, but, as we can see, the corrections for the solution $v_1=v_2\equiv v$, i.e., when the radius of $AdS_2$ is equal to the one of the $S^{2}$, 
are zero. However, there are still other solutions that need to be analyzed. For $v_1\neq v_2 $ the equations
\be 
\left(\frac{e^{2}}{v_{1}^{2}}+\frac{P^{2}}{v_{2}^{2}}\right) +2\tilde\alpha \left(\frac{e^{2}}{v_{1}^{3}}-\frac{P^{2}}{v_{2}^{3}}\right)=0, 
\ee
give a relation between $v_{1}$ and $v_{2}$. By subtracting eqs. (\ref{ateq1}) and (\ref{ateq2}), we obtain 
\be 
\frac{8}{16\pi G}-(v_{1}+v_{2})\left(\frac{e^{2}}{{v_1}^{2}}+\frac{P^{2}}{{v_2}^{2}}\right)
-2\tilde\alpha\frac{{v_{2}}}{{v_1}^{3}}e^{2}+2\tilde\alpha\frac{{v_{1}}}{{v_2}^{3}}P^{2}=0. 
\ee
We can combine these last two equations to obtain
\be 
\frac{8}{16\pi G}+2\tilde\alpha\left(\frac{e^{2}}{{v_1}^{2}}-\frac{P^{2}}{{v_2}^{2}}\right)=0. \label{comb}
\ee
Replacing the common term in the entropy function, the result is
\begin{equation}
{\mathcal{E}}=\frac{2\pi v_{2}}{G}.
\end{equation}
This solution has twice the entropy of the $v_1=v_2$ solution (for which ${\cal E}=\pi v_2/G$), at least for $\tilde\a\rightarrow 0$, hence it does not
minimize the entropy function, and needs to be discarded. We expect the same to happen in the case considered in the main text, hence we only 
consider the $v_1=v_2$ case.

\bibliographystyle{utphys}
\bibliography{entropy}

\end{document}